# Do Type Ia Supernovae Explode inside Planetary Nebulae?

Travis Court[1,2] , Carles Badenes[1,2] , Shiu-Hang Lee[3] , Daniel Patnaude[4] , Guillermo García-Segura[5] , and Eduardo Bravo[6]
[1] Department of Physics and Astronomy, University of Pittsburgh, 3941 O'Hara Street, Pittsburgh, PA 15260, USA
[2] Pittsburgh Particle Physics, Astrophysics, and Cosmology Center (PITT PACC), University of Pittsburgh, Pittsburgh, PA 15260, USA
[3] Department of Astronomy, Kyoto University Oiwake-cho, Kitashirakawa, Sakyo-ku, Kyoto 606-8502, Japan
[4] Center for Astrophysics | Harvard & Smithsonian, 60 Garden Street, Cambridge, MA 02138, USA
[5] Instituto de Astronomía, Universidad Nacional Autónoma de México, Km. 107 Carr. Tijuana-Ensenada, 22860, Ensenada, B.C., Mexico
[6] Departamento de Física Teórica y del Cosmos, Universidad de Granada, E-18071 Granada, Spain

## Abstract

The nature of Type Ia supernova (SN Ia) explosions remains an open issue, with several contending progenitor scenarios actively being considered. One such scenario involves an SN Ia explosion inside a planetary nebula (PN) in the aftermath of a stellar merger triggered by a common envelope (CE) episode. We examine this scenario using hydrodynamic and nonequilibrium ionization simulations of the interaction between the SN ejecta and the PN cocoon into the supernova remnant (SNR) phase, focusing on the impact of the delay between the CE episode and the SN explosion. We compare the bulk dynamics and X-ray spectra of our simulated SNRs to the observed properties of known Type Ia SNRs in the Milky Way and the Magellanic Clouds. We conclude that models where the SN explosion happens in the immediate aftermath of the CE episode (with a delay $\lesssim 1000$ yr) are hard to reconcile with the observations, because the interaction with the dense PN cocoon results in ionization timescales much higher than those found in any known Type Ia SNR. Models with a longer delay between the CE episode and the SN explosion ($\sim$10,000 yr) are closer to the observations, and may be able to explain the bulk properties of some Type Ia SNRs.

*Unified Astronomy Thesaurus concepts:* Supernova remnants (1667); Type Ia supernovae (1728); Planetary nebulae (1249); Common envelope evolution (2154); X-ray astronomy (1810)

## 1. Introduction

Despite the fact that Type Ia supernovae (SNe Ia) remain a foundational tool in modern cosmology, the precise nature of their progenitor systems remains unknown. Contending theoretical scenarios provide different explanations for the process whereby a carbon-oxygen white dwarf (WD) in a binary system becomes unstable, undergoes a thermonuclear runaway, and is completely disrupted in an SN Ia explosion. In the single degenerate (SD) scenario, this process involves the gradual growth of the WD through accretion of material from a nondegenerate companion. In the double degenerate (DD) scenario, the WD undergoes a more rapid destabilization as a result of a merger or a collision with a degenerate companion (see Wang & Han 2012; Maoz et al. 2014; Seitenzahl & Townsley 2017; Liu et al. 2023, for reviews). Both these scenarios require orbital separations that are orders of magnitude smaller than is possible for main-sequence binaries, and which can only be achieved through common envelope (CE) evolution (Ivanova et al. 2013). While the need for at least one CE episode in the evolutionary pathways of all SN Ia progenitors is well established, there is no consensus about how many episodes are necessary, exactly when they take place, or what is the time elapsed between the last CE episode and the SN explosion (Ruiter et al. 2009; Claeys et al. 2014; Meng & Podsiadlowski 2017).

A series of theoretical studies (Kashi & Soker 2011; Ilkov & Soker 2012, 2013; Tsebrenko & Soker 2015a; Aznar-Siguán et al. 2015) proposed that the SN Ia explosion can happen in the immediate aftermath of a CE episode. In this framework, a red giant branch (RGB) or an asymptotic giant branch (AGB) star ejects part of its envelope mass through dynamical interaction with a WD companion in a classic CE episode, leading to a substantial loss of angular momentum and a drastic reduction of the orbital period of the system (Ricker & Taam 2012; Ohlmann et al. 2016). The traditional view was that the short-period central binary emerging from the CE episode would take a long time to merge (e.g., Taam et al. 1978; Meyer & Meyer-Hofmeister 1979), but Kashi & Soker (2011) pointed out that a sizeable fraction of the envelope mass might not be ejected from the system, falling back and forming a circumbinary disk. The interaction with this disk can then cause the WD and the leftover stellar core to plunge toward each other, triggering a merger or a collision and leading to an SN Ia explosion a short time after the CE episode.

A direct prediction of this scenario is that the material ejected in the CE episode should still be in the vicinity of the progenitor when the SN explodes. There is some uncertainty about the morphology and kinematics of the material ejected during the CE episode, but several observational and theoretical studies have established that a large fraction of bipolar planetary nebulae (PNe) are likely the result of CE evolution (Paczynski 1976; Livio & Soker 1988; Soker & Livio 1994; Nordhaus & Blackman 2006; De Marco 2009; Jones 2017; Ondratschek et al. 2022). Due to the PN-CE association, this scenario is sometimes referred to as a "supernova inside a planetary nebula" (SNIP; Tsebrenko & Soker 2013). The SNIPs framework has been invoked as an explanation for the presence of variable absorption features in the high-resolution spectra of some SN Ia (Sternberg et al. 2011; Tsebrenko & Soker 2015b; Graham et al. 2015; see also Cikota et al. 2017), as well as the morphological features of some young Type Ia supernova remnants (SNRs), like the "ears" seen in Kepler, G299.2−2.9, or G1.9+0.3 (Tsebrenko & Soker 2013, 2015b; Chiotellis et al. 2020, 2021), but the full implications for the dynamical and spectral properties of Type Ia SNRs have never been worked out in detail. This is the goal of the present study.

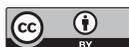

Table 1
Planetary Nebula Cocoon Models from García-Segura et al. (2018)

| Model Name | PN Model | Type | $M_{\mathrm{ZAMS}}(M_\odot)$ | $M_{\mathrm{Core}}(M_\odot)$ | Pre-CE $\dot{M}$ ($M_\odot$ yr$^{-1}$) | $T_{\mathrm{eff}}$ (K) | PN Age (yr) |
|---|---|---|---|---|---|---|---|
| Young, High $\dot{M}$ | A4 | TP-AGB | 1 | 0.569 | $10^{-6}$ | 55,000 | 1000 |
| Old, High $\dot{M}$ | A4 | TP-AGB | 1 | 0.569 | $10^{-6}$ | 55,000 | 10,000 |
| Young, Low $\dot{M}$ | A1 | RGB | 1 | 0.569 | $10^{-8}$ | 29,000 | 1000 |
| Old, Low $\dot{M}$ | A1 | RGB | 1 | 0.569 | $10^{-8}$ | 29,000 | 10,000 |

This paper is organized as follows. The CE cocoon ambient medium (AM) models representative of our PNe are described in Section 2.1. The explosion models and simulation code are described in Section 2.2. In Section 3, we compare the bulk properties of our models to the observed SNRs. Lastly, in Section 4 we summarize our results and outline future study with this framework.

## 2. Methods

### 2.1. Planetary Nebula Models

Several theoretical studies have simulated the formation of PN-like cocoons during CE episodes (Nordhaus & Blackman 2006; García-Segura et al. 2018, 2020, 2021, 2022; Ondratschek et al. 2022). The specific details vary, but all simulations feature the ejection of slow, dense material along the orbital plane of the binary and faster, lower-density material along the perpendicular direction, leading to a more or less collimated bipolar structure. To evaluate the SNIPs scenario, we choose two representative simulations from the study by García-Segura et al. (2018). These models use high-resolution CE calculations from Ricker & Taam (2012) as initial conditions, and follow the 2D hydrodynamic and ionization evolution of the PN cocoon under a wide range of physical assumptions (see Table 1 in García-Segura et al.2018 and the accompanying text for descriptions of all models). We will focus on the A1 and A4 simulations. Model A1 is an AGB star with a zero-age main sequence (ZAMS) mass of 1 $M_\odot$, a core mass of 0.57 $M_\odot$, an initial core temperature of $T_{\mathrm{eff}} = 29,000$ K, and a mass-loss rate prior to the CE of $10^{-8} M_\odot$ yr$^{-1}$. Model A4 is a thermally pulsating AGB star with the same ZAMS and core mass, but a larger initial core temperature ($T_{\mathrm{eff}} = 55,000$ K) and higher mass-loss rate prior to the CE ($10^{-6} M_\odot$ yr$^{-1}$). This information is tabulated in Table 1.

The basic properties and temporal evolution of the PNe sculpted by models A1 and A4 are discussed in detail in García-Segura et al. (2018). For our purposes of evaluating the SNIPs scenario, we will select an early snapshot of each model, 1000 yr after the CE ejection (henceforth the "Young PN" models), and a later snapshot of each model 10,000 yr after the CE ejection (henceforth the "Old PN" models). Thus, we end up with four different PN profiles: young, high-$\dot{M}$ (A4, 1000 yr); young, low-$\dot{M}$ (A1, 1000 yr); old, high-$\dot{M}$ (A4, 10,000 yr); and old, low-$\dot{M}$ (A1, 10,000 yr). One-dimensional density profiles for these four profiles along the equatorial and

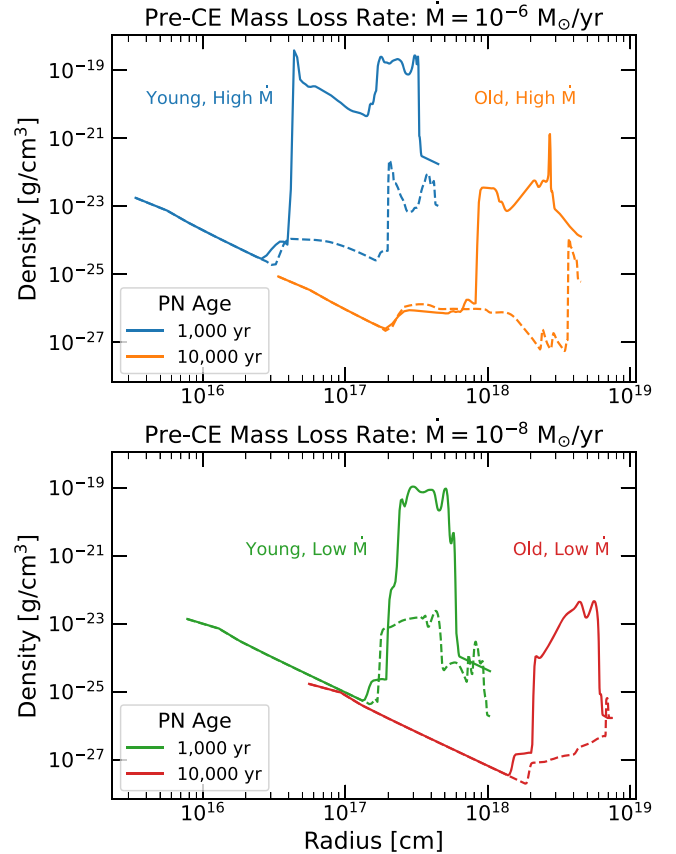

**Figure 1.** One-dimensional density profiles of PN models evolved from CE simulations, shown as equatorial (solid) and polar (dashed) projections of the 2D models from García-Segura et al. (2018). Top: the blue and orange lines correspond to the higher pre-CE mass-loss rate of $\dot{M} = 10^{-6} M_\odot$ yr$^{-1}$ (i.e., Model A4) at a PN age of 1000 yr and at an age of 10,000 yr respectively. Bottom: the green and red lines correspond to the lower pre-CE mass-loss rate of $\dot{M} = 10^{-8} M_\odot$ yr$^{-1}$ (i.e., Model A1) at a PN age of 1000 and 10,000 yr (see García-Segura et al. 2018 for details).

polar directions are shown in Figure 1. These models cover a large dynamic range in AM density and spatial distribution, but they have some common features. Among these are a freely expanding $\rho \propto r^{-2}$ outflow region close to the inner binary, surrounded by denser material, which is always densest along the equatorial direction, sometimes by as much as 5 orders of magnitude. The outer edge of the models is smoothly joined to a uniform density of $\rho = 1.0 \times 10^{-24}$ g cm$^{-3}$ in all cases. From Ricker & Taam (2012), the ejected envelope mass is 0.172 $M_\odot$ and the bound envelope mass that goes on to form the PN cocoon is 0.517 $M_\odot$ (and see Table 2 in García-Segura et al. 2020 for summary). The time elapsed since the CE episode has a strong impact on the density of the AM material and its proximity to the inner binary, with longer delay times resulting in larger, more diffuse cocoons. We note that in the scenario laid out by Kashi & Soker (2011), this delay time can be much

**Table 2**
Summary of Type Ia Explosion Models (see Bravo et al. 2019 for Details)

| Explosion Model | $M_{\rm WD}$ ($M_\odot$) | $\rho_c$ (g cm$^{-3}$) | $\rho_{\rm DDT}$ (g cm$^{-3}$) | $E_K$ (erg) |
|---|---|---|---|---|
| sch088 | 0.88 | $1.5 \times 10^7$ | N/A | $0.917 \times 10^{51}$ |
| sch115 | 1.15 | $9.5 \times 10^7$ | N/A | $1.460 \times 10^{51}$ |
| ddt12 | 1.4 | $3.0 \times 10^9$ | $1.2 \times 10^7$ | $1.182 \times 10^{51}$ |
| ddt40 | 1.4 | $3.0 \times 10^9$ | $4.0 \times 10^7$ | $1.493 \times 10^{51}$ |

**Note.** All models have a progenitor metallicity of $Z = 9 \times 10^{-3}$ and a temperature of $T_{\rm WD} = 1 \times 10^8$ K.

shorter than the 1000 yr we have chosen for the Young PN models. We will return to the issue of the time elapsed since the CE episode in Section 3. We note that the PN models of García-Segura et al. (2018) were not produced with the specific goal to model Type Ia SN progenitors. To our knowledge, no such detailed calculations exist. Yet the models presented here are representative of the class of low-mass post-CE cocoons that is relevant to the SNIPs scenario in particular.

### 2.2. Type Ia Explosion Models

Two broad classes of explosion models have shown success in reproducing the main observational properties (light curves and spectra) of Type Ia SNe (Maoz et al. 2014; Seitenzahl & Townsley 2017). The first of these classes involves a CO WD with a mass close to the Chandrasekhar limit ($M_{\rm Ch} = 1.4\ M_\odot$), where a deflagration burning front starts to propagate through the inner regions of the WD and then transitions to a detonation once it reaches a specific density (Khokhlov 1991). The other class involves the detonation of a CO WD that is below the $M_{\rm Ch}$ (Sim et al. 2010). We will use four representative models from these two classes published by Bravo et al. (2019), selected to bracket the extremes of the parameter space that shows best agreement with Type Ia SN observations. The details of these four models are presented in Table 2.

### 2.3. Supernova Remnant Models and Synthetic X-Ray Spectra

We conduct 1D simulations of the interaction between the PN cocoons described in Section 2.1 and the SN Ia explosion models described in Section 2.2 using ChN, a multipurpose code that couples hydrodynamics (HD), nonequilibrium ionization (NEI), plasma emissivities, radiative cooling, and forbidden line emission (Ellison et al. 2007, 2010; Patnaude et al. 2010; Lee et al. 2012, 2013, 2014, 2015; Patnaude et al. 2017; Martínez-Rodríguez et al. 2018; Jacovich et al. 2021).

To accurately model the interaction with the dense AM in the PN cocoons without running into time-step issues due to the Courant condition, we have developed a Eulerian version of ChN. To initiate the simulations in this Eulerian grid, we expand all ejecta profiles homologously to a time between 0.1 and 0.3 yr after the explosion. This prevents large density contrasts between the outermost ejecta and the innermost AM layers, which would compromise convergence, but does not affect the final outcome (Dwarkadas & Chevalier 1998; Badenes et al. 2003). The Eulerian grid resolution is enhanced in this innermost region to better model the initial stages of the interaction. Elsewhere, the grid is constructed with a mixture of logarithmic and linear spacing to resolve the local structures in the inner PN cocoon without using too many layers in the outer regions.

The physics included in the code remains unchanged from the Lagrangian version, except for the fact that the HD equations are solved on a fixed Eulerian grid, and that physical processes occurring locally in the plasma rest frame, such as the equilibration of electron and ion temperatures and the evolution of the ion fractions, are followed by Lagrangian tracer particles. This approximation is valid for these kinds of processes, which do not provide feedback to the HD. In our simulations, we use 500 tracer particles, which are inserted on the grid at the start of the run and advected in a comoving fashion, following the HD velocity field after each hydro time step. We have performed control runs with a higher number of tracer particles to ensure that the results converge. A fourth-order Runge–Kutta method is used to interpolate the HD quantities for each tracer particle after the update of its spatial location. A similar approach has been implemented to follow explosive nucleosynthesis in three-dimensional codes such as FLASH (Fryxell et al. 2000). Our convergence tests ensure that the benefit obtained from the Eulerian HD in terms of computational cost and the ability to model denser AM interaction does not result in any negative trade-offs on the precision of the thermal physics calculations or the properties of the synthetic spectra obtained from the models.

Because our SNR simulations are 1D, we take the density profiles along the equatorial and polar directions shown in Figure 1 as limiting cases, bracketing the dynamic range of AM densities in each PN model. This approach, sometimes known as the radial flow approximation (Chevalier & Soker 1989), allows us to evaluate the bulk dynamics and ionization timescales of our SNR models in a way that is simplified, but accurate enough for a realistic comparison to spatially integrated observations of real objects. Of course, our 1D simulations cannot capture all the features of the SNIPs scenario, but they are realistic enough to single out the areas of parameter space that show more promise and should be examined using multi-D calculations in future work. We evolve our models to an age of 5000 yr after the SN explosion.

For each simulation, we calculate synthetic X-ray spectra using the NEISession package of the pyatomdb module (Foster & Heuer 2020), which we feed with the output from ChN: the electron and ion temperatures, density, chemical composition, and ionization state for each location in the SNR model. The raw synthetic spectra produced in this way can be convolved with specific instrumental responses in order to compare to observations. In Figure 2, we show the temporal evolution of the X-ray spectra emitted by the shocked ejecta in the SNR resulting from the interaction between the near-$M_{\rm Ch}$ explosion model ddt40 and the equatorial projection of the old, low-$\dot{M}$ PN model, convolved with both the Suzaku and XRISM spectral responses. In this and other figures, we do not show the thermal emission from the shocked AM, which usually makes a small contribution to the spatially integrated X-ray spectra of young Type Ia SNRs (e.g., Decourchelle et al. 2000, 2001; Badenes et al. 2008). Specifically, we note that adding the contribution from the shocked AM does not have a strong impact on the properties of the line emission from high-$Z$ elements like Si and Fe, which are the focus of our analysis. The spectra in Figure 2 showcase the temporal evolution of the ionization timescale of the X-ray-emitting plasma—note, for example, the increase in the centroid energy of the Fe K$\alpha$ line blend at $\sim$6.5 keV and the rise of the Si Ly$\alpha$ to Si K$\alpha$ line ratio as the mean charge state of Si approaches the H-like ion (Si XIV). By the age of 110 yr, the forward shock in this SNR model has entered the densest part of the equatorial profile, shown in the bottom panel of Figure 1 in red. After 460 yr, the forward shock has propagated beyond the PN cocoon and is expanding into the uniform AM outside of the PN structure.

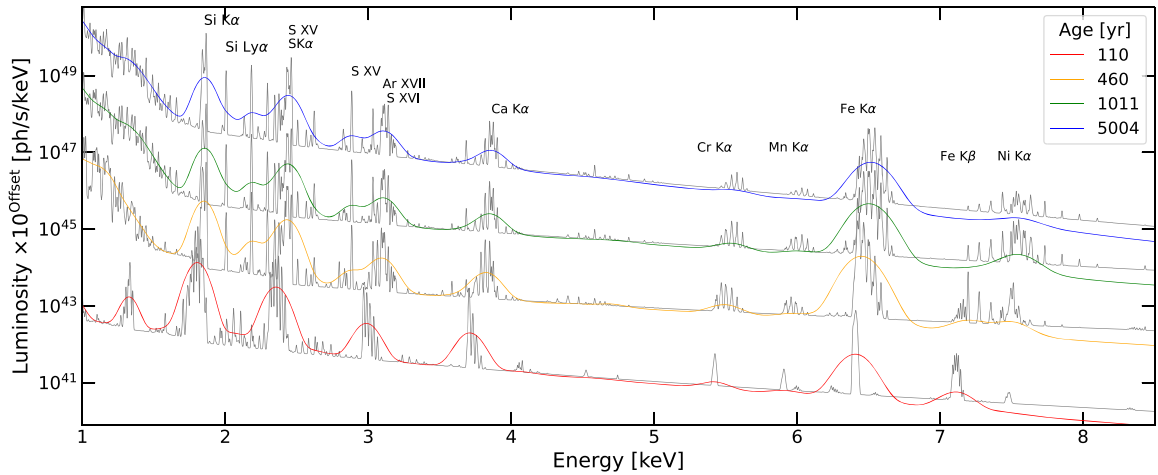

**Figure 2.** Reverse shock spectral evolution for `ddt40` explosion of the equatorial projection of the $\dot{M} = 10^{-6}\,M_\odot\,\mathrm{yr}^{-1}$ PN at an age of 10,000 yr. The gray spectra are convolved with a XRISM response matrix, while spectra with color are convolved with a Suzaku response matrix. The luminosity of the spectra is offset by an arbitrary amount to show the relative feature strength and evolution of features.

## 3. Comparison to Observations: Evaluating the SNIPs Scenario

Although our HD+NEI SNR models can be used for comparison to individual objects (see Badenes et al. 2006, 2008; Patnaude et al. 2012, for examples), here we will take a more general approach to evaluate the SNIPs scenario. We will focus on the centroid energy of the Fe K$\alpha$ blend, an excellent tracer of the mean ionization timescale of the shocked Fe-rich plasma, which is dominated by SN ejecta in young SNRs. Yamaguchi et al. (2014a) noticed a stark dichotomy in a sample of 23 young SNRs in the Large Magellanic Cloud (LMC) and the Milky Way with Fe K$\alpha$ emission observed by Suzaku and Chandra. All SNRs known or suspected to be of Type Ia origin showed Fe K$\alpha$ centroid energies below 6.55 keV (roughly equivalent to the Fe XXII ion; Yamaguchi et al. 2014b), while those known or suspected to be of core-collapse (CC) origin showed higher centroid energies, and therefore more ionized Fe. The most straightforward interpretation of this result is that the AM around Ia SNe is less dense than around CC SNe, leading to longer timescales for ionization equilibration and a lower degree of ionization at similar SNR ages. This result has been confirmed by subsequent studies (Maggi et al. 2016), and although there might be some exceptions (e.g., the likely CC SNR G15.9+0.2, which shows a low Fe K$\alpha$ centroid; Maggi & Acero 2017), it is clear that any SNR model that results in Fe K$\alpha$ centroid energies higher than 6.55 keV will not be able to match the observational properties of most known Type Ia SNRs.

Previous work (Yamaguchi et al. 2014a; Martínez-Rodríguez et al. 2018) has shown that the interaction of SN Ia explosion models with a uniform AM of a density typical of the warm phase of the interstellar medium (ISM) (e.g., $\rho_{AM}$ between 0.05 and $5 \times 10^{-24}\,\mathrm{g\,cm}^{-3}$) is fairly successful at explaining the fundamental observational properties (Fe K$\alpha$ centroid and bulk dynamics) of young Type Ia SNRs as a class. This does not mean that the AM around all SN Ia progenitors is uniform—indeed, objects like Kepler (Reynolds et al. 2007; Patnaude et al. 2012), N103B (Williams et al. 2014; Yamaguchi et al. 2021) and RCW 86 (Williams et al. 2011; Patnaude & Badenes 2017) are hard to reconcile with a uniform AM interaction. But it does put strong constraints on the density and location of the AM in the vicinity of SN Ia progenitors, and therefore on their pre-SN mass-loss rate (see Badenes et al. 2007; Patnaude & Badenes 2017, for discussions).

To illustrate these limits in the context of the SNIPs scenario, we show in Figures 3 and 4 the luminosity of the Fe K$\alpha$ line, SNR radius, and SNR age as a function of the Fe K$\alpha$ line centroid in SNRs produced from the interaction between the `ddt40` SN Ia ejecta and our PN models, together with a compilation of observed values for Type Ia and CC SNRs in the Milky Way and the LMC (from Yamaguchi et al. 2014a; Martínez-Rodríguez et al. 2018). The Fe K$\alpha$ centroids and luminosities shown in the plots are calculated from raw unconvolved model spectra, but the results are very similar if the models are convolved with a spectral response; see the Appendix for details. These plots are similar to Figure 13 in Martínez-Rodríguez et al. (2018), which shows the same comparison between SNR models and observations for the case of an interaction with a uniform AM. We show SNR models calculated using both the equatorial (thick lines) and the polar (thin lines) profiles of each PN cocoon to bracket the extremes of variation. However, because the emission measure of hot, X-ray-bright plasmas scales with the square of the density, the predictions from the equatorial (i.e., denser) PN profiles will be more representative, and closer to the values that would be measured in a spatially integrated observation.

In Figure 3, we show the interaction between the `ddt40` explosion model and the young PN profiles (left panel for the $\dot{M} = 10^{-6}\,M_\odot\,\mathrm{yr}^{-1}$ model, right panel for the $\dot{M} = 10^{-8}\,M_\odot\,\mathrm{yr}^{-1}$ model). The high-density material found along the equatorial plane of these young PN models is clearly incompatible with the observational properties of Type Ia SNRs. Strong AM interaction early in the evolution of these SNR models quickly drives the ionization timescale of the shocked plasma to extreme values, with Fe K$\alpha$ centroid energies in excess of 6.65 keV, roughly corresponding to the Fe XXIV ion. These centroid energies are comparable to those found in CC SNRs interacting with very dense AM like W49B, which has a strong radiative recombination continuum (Ozawa et al. 2009). Indeed, both SNR models go through a recombination phase (i.e., the Fe K$\alpha$ centroid evolves toward lower energies) for the first few hundred years. Afterwards, the SNRs expand and become fainter in the Fe K$\alpha$ line, but the ionization timescales remain high throughout the evolution. These SNR models also predict higher Fe K$\alpha$ luminosities than are found in any SNR, regardless of type. The interaction with the polar profiles of these young PN models shows better agreement with the observations, but this is not significant, because the emission-measure effect discussed in the previous paragraph would heavily bias any spatially integrated measurements toward the predictions of the equatorial profiles. To put this in perspective, the total emission measure of the SNR model corresponding to the equatorial profile of the young, high-$\dot{M}$ PN model is a factor $\sim 40$ higher than the one corresponding to the polar profile at an age of 550 yr.

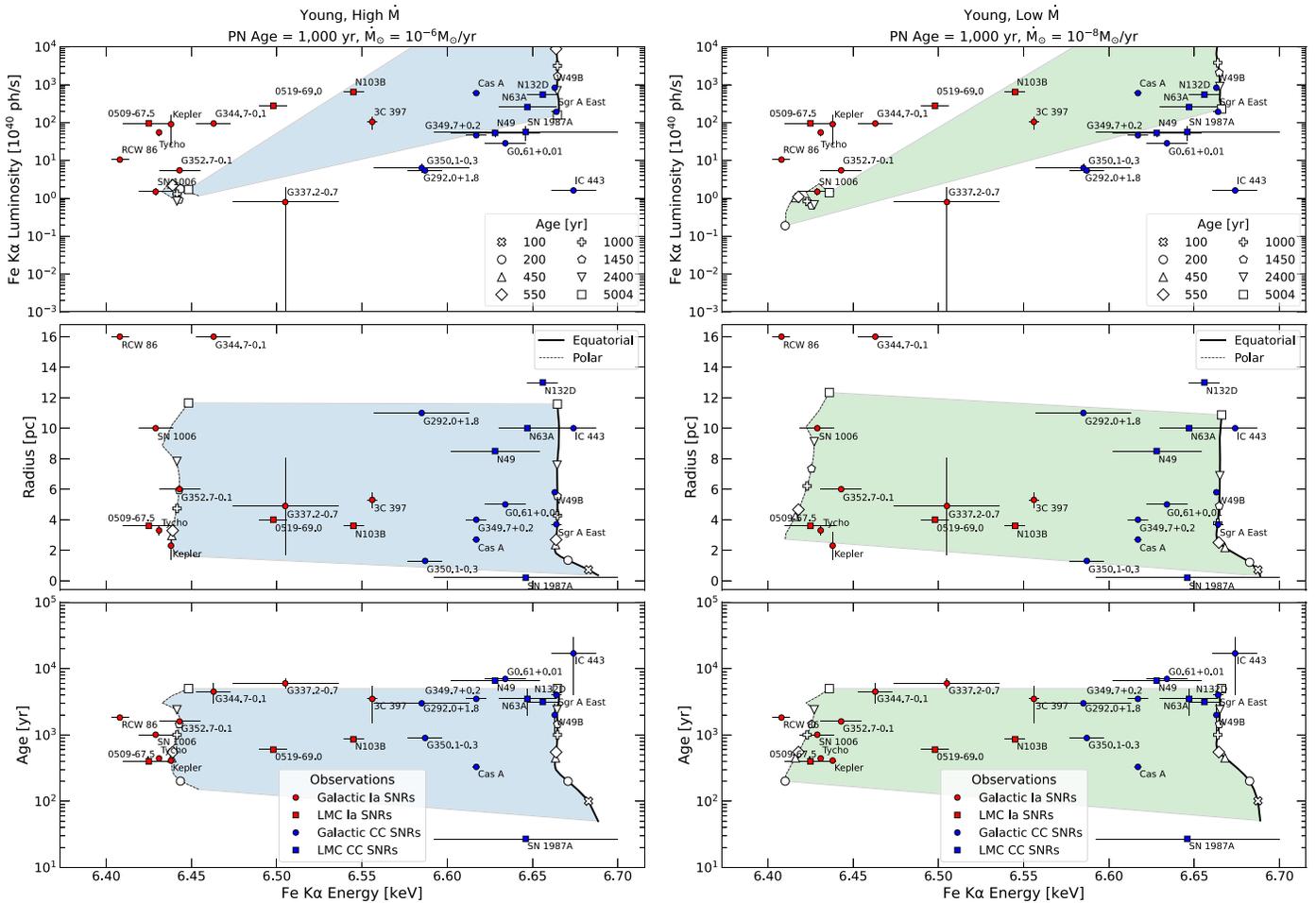

**Figure 3.** Fe K$\alpha$ luminosity, SNR radius, and SNR age as a function of Fe K$\alpha$ centroid energy for the `ddt40` ejecta interacting with the high mass-loss rate (left, blue) and low mass-loss rate (right, green) PN models at an age of 1000 yr. The red and blue symbols represent measurements of known Type Ia and CC SNRs, respectively, from Yamaguchi et al. (2014a) and Martínez-Rodríguez et al. (2018). For each PN model, we show both the equatorial (thick line) and polar (thin line) 1D profiles, highlighting the region between them with a shade, but we remind the reader that the denser (equatorial) profile will be closer to any spatially integrated SNR measurements like the ones shown here (see text for details).

The interaction with the old PN cocoons (10,000 yr after CE ejection), shown in Figure 4, tells a different story. In the SNR model corresponding to the interaction between the SN ejecta and the equatorial profile of the old, high-$\dot{M}$ PN (left panel, thick line), the shocked ejecta steadily increases its ionization timescale up to a sharp peak at an SNR age of 1000 yr, produced by a secondary shock reflected on the density peak located at $\sim 3 \times 10^{18}$ cm in Figure 1. After this sharp peak, the plasma recombines and the Fe K$\alpha$ line luminosity decreases, but the model remains solidly in the area of parameter space populated by CC SNRs. By contrast, the SNR model corresponding to the equatorial profile of the old, low-$\dot{M}$ PN, shown in the right panel, predicts Fe K$\alpha$ centroids and luminosities that are quite close to those of known Type Ia SNRs. The evolution of this SNR model is markedly different from the young PN cases: The Fe K$\alpha$ centroid increases gradually with age, while the Fe K$\alpha$ luminosity first increases, as the forward shock plunges into the dense part of the cocoon, plateaus for a while, and then becomes fainter as the SNR expands and the forward shock propagates beyond the cocoon. Without making an attempt to model any specific objects, we note that this old, low-$\dot{M}$ PN model is quite close to the bulk properties of the LMC SNR 0519−69.0 at an SNR age between 550 and 900 yr, which is not far from its estimated age of $\sim 600$ yr (Kosenko et al. 2010).

Not surprisingly, our SNR models evolving into PN cocoons are not a good match for physically large SNRs like RCW 86, which is likely an SN Ia explosion evolving into a large, low-density cavity excavated by a fast outflow from the progenitor (Badenes et al. 2007; Williams et al. 2011; Patnaude & Badenes 2017). The PN cocoons we consider here are excavated by slow outflows (Koo & McKee 1992a, 1992b), leading to momentum-driven structures with relatively high densities that contain the expansion of the SNR to some extent. This results in objects that are physically smaller at the same age than SNRs expanding into a uniform AM. To put this comparison in context, we include an SNR model obtained from the interaction between the ddt40 ejecta and a uniform AM of density $1.0 \times 10^{-24}$ g cm$^{-3}$ in Figure 4 as a thick gray line. For the same SNR age, this uniform AM produces SNRs with lower Fe K$\alpha$ centroid energies, lower Fe K$\alpha$ luminosities, and larger radii than the old, low-$\dot{M}$ PN.

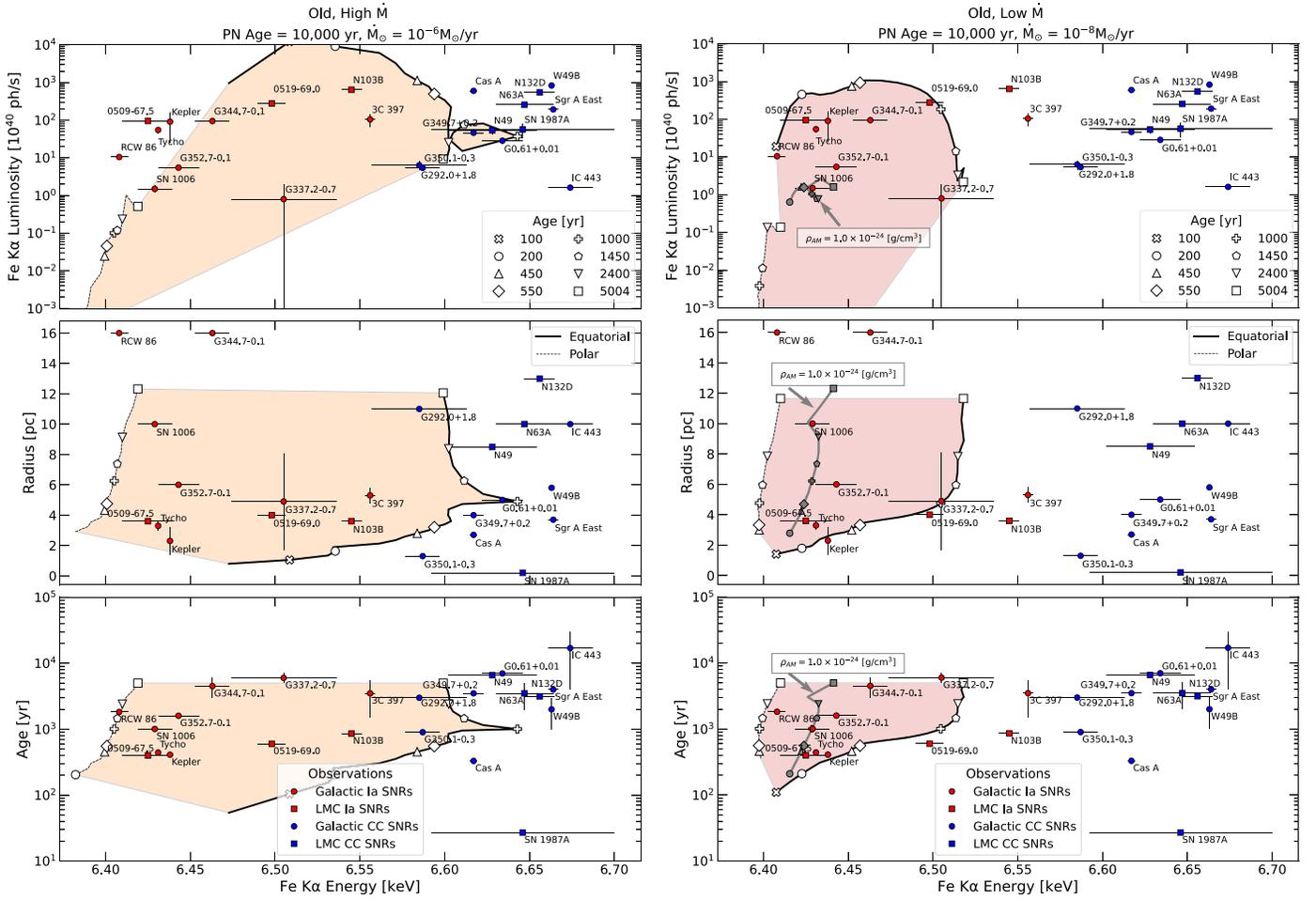

**Figure 4.** Fe Kα luminosity, SNR radius, and SNR age as a function of Fe Kα centroid energy for the `ddt40` ejecta interacting with the high mass-loss rate (left, orange) and low mass-loss rate (right, red) PN models at an age of 10,000 yr. The red and blue symbols represent measurements of known Type Ia and CC SNRs, respectively, from Yamaguchi et al. (2014a) and Martínez-Rodríguez et al. (2018). For each PN model, we show both the equatorial (thick line) and polar (thin line) 1D profiles, highlighting the region between them with a shade, but we remind the reader that the denser (equatorial) profile will be closer to any spatially integrated SNR measurements like the ones shown here (see text for details). For comparison, we add an SNR model obtained assuming a uniform AM with a density of $1.0 \times 10^{-24}$ g cm$^{-3}$ in gray (see text for details).

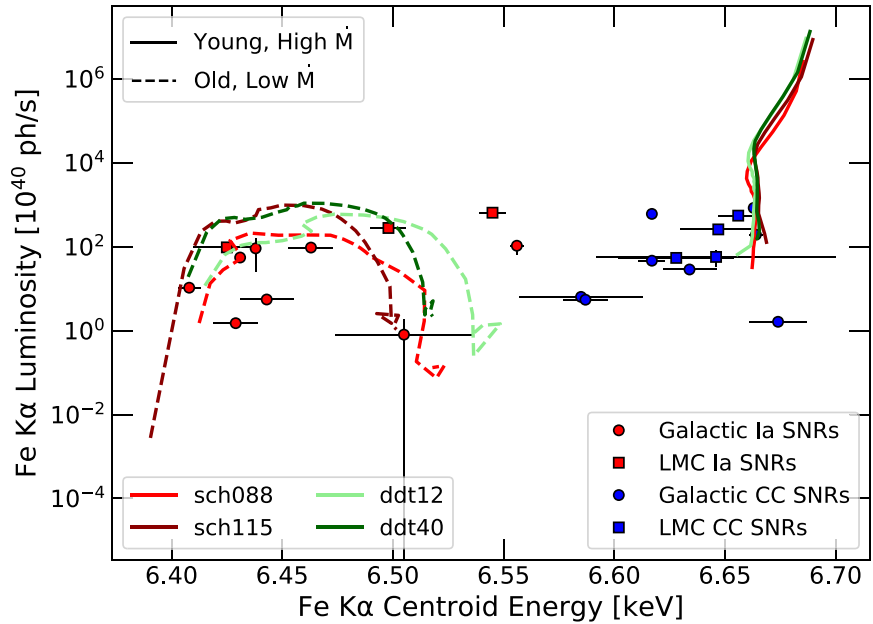

**Figure 5.** Fe Kα luminosity and centroid energy comparison for different SNe Ia explosion models. Each colored line represents a different explosion model used with the PN cocoons. For clarity, only the 1D equatorial projections of the PNe are shown. The younger PN model with the higher mass-loss rate is shown with solid lines, while the older PN model with lower mass-loss rate is shown with dashed lines. The colors of the lines each represent a different SN Ia explosion model, with both sub-$M_{\rm Ch}$ models and delayed-detonation near-$M_{\rm Ch}$ models.

In Figure 5, we show the effect of varying the SN Ia explosion model for the two PN models on the farthest reaches of AM density parameter space. As noted by Patnaude et al. (2015) and Martínez-Rodríguez et al. (2018), the bulk properties of young SNRs are driven mainly by the strength of the AM interaction. Therefore, our broad conclusions regarding the viability of the PN models are not sensitive to the properties of SN Ia explosions. This would not hold true for detailed comparisons with individual objects, which do have the potential to distinguish between different explosion models (Badenes et al. 2006, 2008; Patnaude et al. 2012), but we leave those comparisons to future work.

## 4. Discussion and Conclusions

We have calculated HD+NEI SNR models resulting from the interaction between Type Ia SN ejecta and four representative post-CE PN cocoons at different ages, and compared them to the bulk properties of known Type Ia SNRs in the Milky Way and the Magellanic Clouds. Our results indicate that it is possible for this sort of interaction to produce models that are close to the properties of observed Type Ia SNRs, but only if there is a substantial time delay between the CE episode and the SN explosion, and the mass-loss rate before the CE is relatively low ($\sim 10^{-8} M_\odot$ yr$^{-1}$). Models where this time delay is only 1000 yr result in AM structures that are quite dense ($\rho \sim 10^{-20}$ g cm$^{-3}$, or $n \sim 4 \times 10^3$ cm$^{-3}$) and very close to the progenitor ($\sim 0.1$ pc). Any SNR expanding into this sort of AM will evolve to be extremely luminous and highly ionized, more similar to CC SNRs known to be expanding in dense media than to Type Ia objects, even the ones that are suspected to be interacting with a nonuniform AM. By contrast, models where the time elapsed between the CE episode and the SN explosion is an order of magnitude larger (10,000 yr) result in less dense AM ($\rho \sim 10^{-23}$ g cm$^{-3}$, or $n \sim 4$ cm$^{-3}$), farther away from the progenitor ($\sim 1$ pc). SNR models expanding into these AM structures can show substantial overlap with the observations of known Type Ia SNRs, provided that the mass-loss rate before the CE event is relatively low. In some cases, like the LMC SNR 0519 −69.0, the old, low-$\dot{M}$ CE cocoon model matches the bulk properties of the SNR quite well. Incidentally, the CE to explosion delay times of 10,000 yr that provide the best match to the SNR observations are comparable to the kinetic ages of many PNe (González-Santamaría et al. 2021), and to the lifetimes of PNe in the LMC derived by statistical analysis (Badenes et al. 2015), but without a detailed understanding of PN progenitors it is hard to interpret this coincidence in a meaningful way.

Our constraints on the time delay between CE ejection and SN explosion are important in the context of the SN Ia progenitor problem. In the scenario laid out by Tsebrenko & Soker (2013), this time delay can be quite small. These authors propose that the interaction with the circumbinary disk can lead to a rapid plunge and a merger or a collision in a substantial fraction of binaries while the CE episode is still occurring. This possibility is clearly ruled out by our work, as CE cocoons with delay times shorter than 1000 yr would lead to even denser AM closer to the progenitor, and more extreme interaction scenarios, perhaps akin to what is seen in Type IIn SNe like SN2005ip (Smith et al. 2009) or SN 2006gy (Smith & McCray 2007; Jerkstrand et al. 2020), but unlike any known Type Ia SNR. From a theoretical point of view, it is hard to constrain the timescale for the final inspiral of the inner binary without detailed numerical simulations, which do not exist for this specific scenario (but see Tiede et al. 2020, for a case study of inspiral triggered by circumbinary disks in the context of supermassive black holes). More approximated semi-analytical treatments imply that the $\sim 10^4$ yr timescales favored by our analysis are certainly within the allowed range in some cases; see Figure 4 in Metzger (2022) and the accompanying discussion.

Our work showcases the importance of modeling the NEI processes that take place in SNRs before comparing simulations to observations. We contend that the interaction between SN ejecta and the AM structures around Type Ia progenitors cannot be evaluated by focusing only on the morphology of the resulting SNR models, without taking into account the bulk ionization timescales predicted by the interaction. It is entirely possible that a specific AM structure can produce a spatial morphology very similar to that of a given SNR, but grossly misrepresent its spectral properties.

We conclude with a word of caution. The parameter space for the structure of the AM around Type Ia SNe is quite large, particularly given the few constraints that we have on the pre-SN properties of the progenitors. There is a substantial body of observational evidence that disfavors large amounts of dense material close to SN Ia progenitors, from X-ray studies of SNRs like the ones discussed here to X-ray and radio observations conducted after the explosions of nearby SNe (Chomiuk et al. 2012; Margutti et al. 2014; Chomiuk et al. 2016). Taken as a whole, these studies probe spatial scales that range between a few thousand au and several parsecs (Patnaude & Badenes 2017). While a few Type Ia SNe do seem to interact with dense CSM—the so-called Ia-CSM SNe (Silverman et al. 2013; Bochenek et al. 2018)—these objects are rare, and represent a small fraction (a few percent) of the total population of core-normal Type Ia SNe (Dubay et al. 2022). We would not expect any such objects to appear in a sample of a dozen local Type Ia SNRs. Theoretical studies that require any sort of dense, slow outflow from SN Ia progenitors in the phases leading to the explosion should take these constraints into account.

## Acknowledgments

We acknowledge discussions with Lars Bildsten, Paul Ricker, Hiroya Yamaguchi, and Mathieu Renzo. T.A.C. and C.B. acknowledge support from Chandra Theory grant Nos. TM0-21004X and TM1-22004X. E.B. acknowledges partial support from the Spanish project PID2021-123110NB-100, financed by MCIN/AEI/10.13039/501100011033/FEDER/UE.

*Software*: ChN (Ellison et al. 2007, 2010; Patnaude et al. 2010; Lee et al. 2012, 2013, 2014, 2015; Patnaude et al. 2017; Martínez-Rodríguez et al. 2018; Jacovich et al. 2021), Numpy (van der Walt et al. 2011), Matplotlib (Hunter 2007), PyAtomDB (https://atomdb.readthedocs.io/en/master/index.html), Pandas (McKinney 2010), SciPy (Virtanen et al. 2020).

## Appendix

We calculate the Fe K$\alpha$ centroid energy, $E_{\mathrm{FeK}\alpha}$, by subtracting the continuum from the nonconvolved spectrum following Martínez-Rodríguez et al. (2018). We separate the continuum by calculating the derivative of the flux with respect to photon energy and then using a sufficiently high threshold to identify spectral lines. Once the continuum is determined in this way, we fit it with a third-degree polynomial and subtract this fit from the flux. This has several advantages. First, it is an automated procedure and yields similar results to those obtained by hand-fitting the polynomial to isolate lines. Second, it is easy to extrapolate to other strong emission lines. Isolating the thermal line emission at different energies is more computationally expensive.

After continuum subtraction, we calculate the centroid energy of the feature,

$$E_{\mathrm{FeK}\alpha} = \frac{\int_{E_{\min}}^{E_{\max}} F \times E\, dE}{\int_{E_{\min}}^{E_{\max}} F\, dE} = \frac{\sum_i F_i \times E_i \times dE_i}{\sum_i F_i \times dE_i}, \quad (A1)$$

where $F$ is the differential flux from the nonconvolved spectrum after continuum subtraction, $dE$ is the constant energy step (1 eV), and $E_{\min} - E_{\max}$ is the energy interval covering the Fe K$\alpha$ complex (6.3–6.9 keV). We only compare these values where the Fe K$\alpha$ emission is clearly greater than the emission from the continuum determined by the equivalent width of the feature. This method of measuring the centroid energy is advantageous to fitting the synthetic spectra with a power law and a Gaussian as our emission models are more complex.

### ORCID iDs

Travis Court https://orcid.org/0000-0003-3837-7201
Carles Badenes https://orcid.org/0000-0003-3494-343X
Shiu-Hang Lee https://orcid.org/0000-0002-2899-4241
Daniel Patnaude https://orcid.org/0000-0002-7507-8115
Guillermo García-Segura https://orcid.org/0000-0002-9899-2292
Eduardo Bravo https://orcid.org/0000-0003-0894-6450